\documentclass[11pt]{article}
\usepackage{xspace}
\usepackage{graphicx}
\usepackage{amsmath}
\usepackage{amssymb}
\usepackage{color}

\definecolor{Red}{rgb}{1,0,0}
\definecolor{Green}{rgb}{0,1,0}
\definecolor{Blue}{rgb}{0,0,1}
\definecolor{Black}{rgb}{0,0,0}

\def\beq{\begin{equation}}
\def\eeq#1{\label{#1}\end{equation}}
\def\eeqn{\end{equation}}

\def\beqa{\begin{eqnarray}}
\def\eeqa#1{\label{#1}\end{eqnarray}}
\def\eeqan{\end{eqnarray}}

\let\bar=\overbar

\def\Dslash{\not{\hbox{\kern-4pt $D$}}}
\def\dslash{\not{\hbox{\kern-2pt $\del$}}}

\def\msb{{\bar{\ssstyle M \kern -1pt S}}}

\def\Title#1{\begin{center} {\Large {\bf #1} } \end{center}}

\begin{document}

\Title{Radiation Background Studies for {0$\nu\beta\beta$} decay in  $^{124}$Sn}

\bigskip\bigskip

%+\addtocontents{toc}{{\it D. Reggiano}}
%+\label{ReggianoStart}

\begin{raggedright}  

%% Authors - you should specify at least one author as follows.
{\it Neha Dokania$^{1,2}$\index{neha}, \it V. Singh$^{1,2}$\index{vivek}, {\it }
\it C. Ghosh$^{3}$ \index{chandan},
\it S. Mathimalar$^{1,2}$\index{mathi},
\it A. Garai$^{1,2}$ \index{garai},
\it S. Pal$^3$ \index{sanjoy},
\it V. Nanal$^3$\index{nanal}, {\it }
\it R.G. Pillay$^3$\index{nanal},
\it{A. Shrivastava$^4$ \index{aradhana}}, K.G. Bhushan$^5$ \index{bhushan}\\
$^1$Homi Bhabha National Institute, Anushaktinagar,
Mumbai.\\
$^2$India-based Neutrino Observatory,
Tata Institute of Fundamental Research,
Mumbai.\\
$^3$Department of Nuclear and Atomic Physics,
Tata Institute of Fundamental Research.\\
$^4$Nuclear Physics Division,
Bhabha Atomic Research Centre,
Mumbai.\\
$^5$Technical Physics Division,
Bhabha Atomic Research Centre,
Mumbai.\\
}

%% In case you want to have more than one author please follow the format
%% shown below, listing the individual authors AND also making sure
%% that each author is given a unique index entry.
%Someone Else\index{Else, S.}, {\it Another University}\\

\end{raggedright}
\vspace{1.cm}

{\small
\begin{flushleft}
\emph{To appear in the proceedings of the Prospects in Neutrino Physics Conference, 15 -- 17 December, 2014, held at Queen Mary University of London, UK.}
\end{flushleft}
}

\section{Introduction}
Neutrinoless Double Beta Decay ($0\nu\beta\beta$) is a lepton number violating process which can probe the nature of neutrino (Majorana or Dirac) and can constrain the absolute neutrino mass scale. A number of experiments using different isotopes and a variety of detection techniques are currently underway to search for $0\nu\beta\beta$ decay. 
In India, a Tin cryogenic bolometer (TIN.TIN) is under development to search for $0\nu\beta\beta$ decay in $\rm^{124}Sn$ and is proposed to be installed at the upcoming India-based Neutrino Observatory (INO)~\cite{inpc}. Given the rarity of the $0\nu\beta\beta$ decay ($\rm T_{1/2}^{0\nu}>10^{24}$ years), ultra-low levels of background $<10^{-2}$ counts/(keV kg day) are required. 
The understanding and minimization of background, especially around the $\rm Q_{\beta\beta}(^{124}Sn)=$ 2292.64$\pm$0.39 keV, is essential to improve the sensitivity of the $ T_{1/2}^{0\nu} $ measurement. This paper describes the low background counting setup at TIFR and the estimated neutron background in the INO cavern.

\section{Tifr Low background Experimental Setup (TiLES)}

A low background counting setup - the TiLES with a high efficiency and low background HPGe detector has been set up at sea level in TIFR. The detector model has been developed using Geant4-based MC simulations for the efficiency determination of different source geometries~\cite{lbkg}.
The detector has graded shield with an inner 5 cm low activity OFHC Cu layer, an outer 10 cm low activity Pb ($^{210}$Pb $<$ 0.3 Bq/kg) layer and an active veto system (plastic scintillators). The sensitivity of the TiLES is measured to be $\sim2$ mBq/g for $^{40}$K and $\sim1$ mBq/g for $^{232}$Th. The setup is used for the qualification and selection of radio-pure materials like TIN.TIN cryostat material ETP Cu, Sn samples and rock samples from the INO site (Bodi West Hills -BWH). The ETP Cu sample showed many gamma rays in the range of 139.4-1326.9 keV formed due to cosmogenic activation while the BWH rock sample had high levels of $^{40}$K (1050 (16) mBq/g)~\cite{lbkg}.

\section{Neutron-induced background in TIN.TIN} 
Neutrons are known to be an important source of background for experiments like direct dark matter searches, double beta decay experiments, etc. The effect of neutron-induced background on the detector and its surrounding materials has been studied using neutron activation technique. The $\rm^{nat,124}$Sn, Teflon, Torlon, $\rm^{ETP}$Cu, $\rm^{nat}$Pb were irradiated at the Pelletron Linac Facility, Mumbai in the neutron irradiation setup with fast neutrons in the energy range of $E_n$ = $\sim$ 0.1 -- $\sim$ 18 MeV. The reaction products formed were mostly short-lived (T$_{1/2}\sim$ 15 min to 80 d). Teflon is found to be better compared to Torlon for support structure in the bolometer, since no high energy gamma activity ($E_{\gamma}>$ 511 keV) was produced due to fast neutron irradiation. Contribution of neutron-induced reactions in different Sn isotopes (A = 112-122) is also investigated. Among the various products of Sn isotopes, $^{123}$Sn has the longest T$_{1/2}$ = 129.2 d, while $\rm^{116m}$In (T$_{1/2}$ = 54.29 m) produces $E_{\gamma}=$ 2112.3 keV near the Q$ _{\beta\beta}$ ($^{124}$Sn) region~\cite{jinst}.

\section{Neutron production in the INO cavern}
In underground laboratories, neutrons originate from the presence of U and Th trace elements in the surrounding rock. The neutrons induced by muon interactions are more penetrating ($E_n > 15$ MeV) but the expected flux is $\sim$ 100--1000 times lower. 
The radioactivity from the rock produces neutrons via spontaneous fission (SF) of U, Th and from the ($\alpha, n$) interactions with the low Z elements present in the rock. 
The BWH rock is mainly Charnockite having a density of $\sim$ 2.89 $\rm g/cm^3$. Its composition is obtained from TOF-SIMS method. Table~\ref{tab:sims} shows various constituent elements of the BWH rock together with respective concentrations. 

\begin{table}[!h]
\begin{center}
\caption[Elemental distributions of BWH rock obtained with TOF-SIMS method]{\label{tab:sims} Elemental distributions of BWH rock obtained with TOF-SIMS method.}
\vspace{5pt}
\begin{tabular}{ l|c || l | c ||l|c  }
\hline\hline
Element&   Concentration  & Element & Concentration & Element & Concentration\\ 
       &  (\% Weight)  &  &  (\% Weight) &  &  (\% Weight)\\ [0.5ex]\hline
$ ^{1} $H, $ ^{2} $H$_2$  & 1.1 & $ ^{6} $Li, $ ^{7} $Li  & 0.001 & $ ^{12} $C& 2 \rule{0pt}{3ex} \\
$ ^{16} $O, $ ^{32} $O$_2$  & 2.04&
$ ^{23} $Na& 5 & $ ^{24,25,26} $Mg& 7 \\
 $ ^{27} $Al & 25 &$ ^{28,29,30} $Si&40&$ ^{31} $P & 3\\
 $ ^{32} $S & 0.1 &
  $ ^{39,40,41} $K & 4 & $ ^{40} $Ca &9.99\\
   $ ^{52} $Cr & 0.49 &
$ ^{56} $Fe  & 1 & $ ^{58,60} $Ni & 0.55 \\
  $ ^{63,65} $Cu & 1.2 &$ ^{107,109} $Ag & 0.05  &$ ^{120} $Sn & 0.6  \\  [0.5ex]  \hline\hline
 \end{tabular}
 \end{center}
\end{table}

\begin{figure}[!ht]
\begin{centering}
\includegraphics[scale=0.5, trim=0.1cm 0cm 0cm 0cm, clip=true]{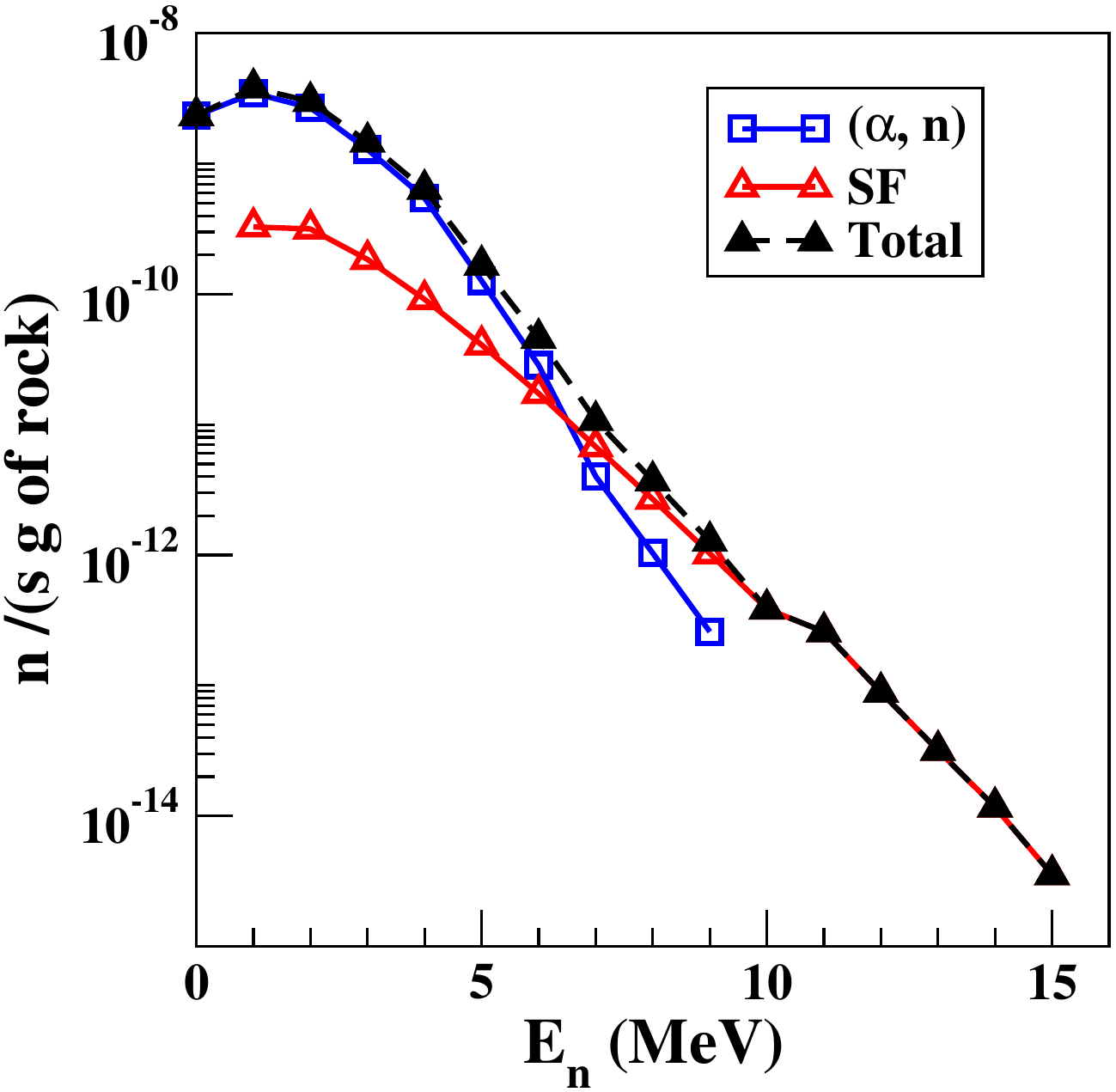} 
\caption[Neutron spectra for BWH rock with 60~ppb of $^{238}$U and 224 ppb of $^{232}$Th]{\label{fig:fiss} Neutron spectra for BWH rock with 60~ppb of $^{238}$U and 224 ppb of $^{232}$Th.}
\end{centering}
\end{figure}

The $^{238}$U and $^{232}$Th content in the BWH rock was obtained using the ICPMS method as 60~ppb and 224~ppb, respectively.
The spectrum of the neutrons emitted in SF is described by an analytic function, known as Watt spectrum, which is given by:
%%%%%%%%%%%%%%%%%%%%%%%%%%%%%%%%%%%%%%%%5
\begin{equation}\label{eqn:watt}
\begin{aligned}
W(a,b,E) &= C e^{-E/a} sinh (\sqrt{b E}),~
{\rm where}~C&= {(\frac{{\pi} b}{4 a})^{\frac{1}{2}}} (\frac{e^{b/4a}}{a})
\end{aligned}
\end{equation}
The thick target ($\alpha$, n) reaction yields for low Z elements (Z$ \leq29 $) have been taken from Refs. \cite{heaton} and \cite{alpha_mei}. The total yield is determined by the sum of the individual element yield weighted by its mass ratio in the BWH rock (as per Table~\ref{tab:sims}). The neutron yields thus obtained, normalized to U and Th content of the BWH rock, are shown in Figure~\ref{fig:fiss}.
It is evident that the ($\alpha, n$) component dominates at lower neutron energies while the SF dominates at higher energies ($E_n > 8$ MeV). 

\section{Neutron flux estimation in the INO cavern}
\begin{figure}[!h]
\begin{center}
\includegraphics[scale=0.5, trim=0.1cm 0cm 0cm 0cm, clip=true]{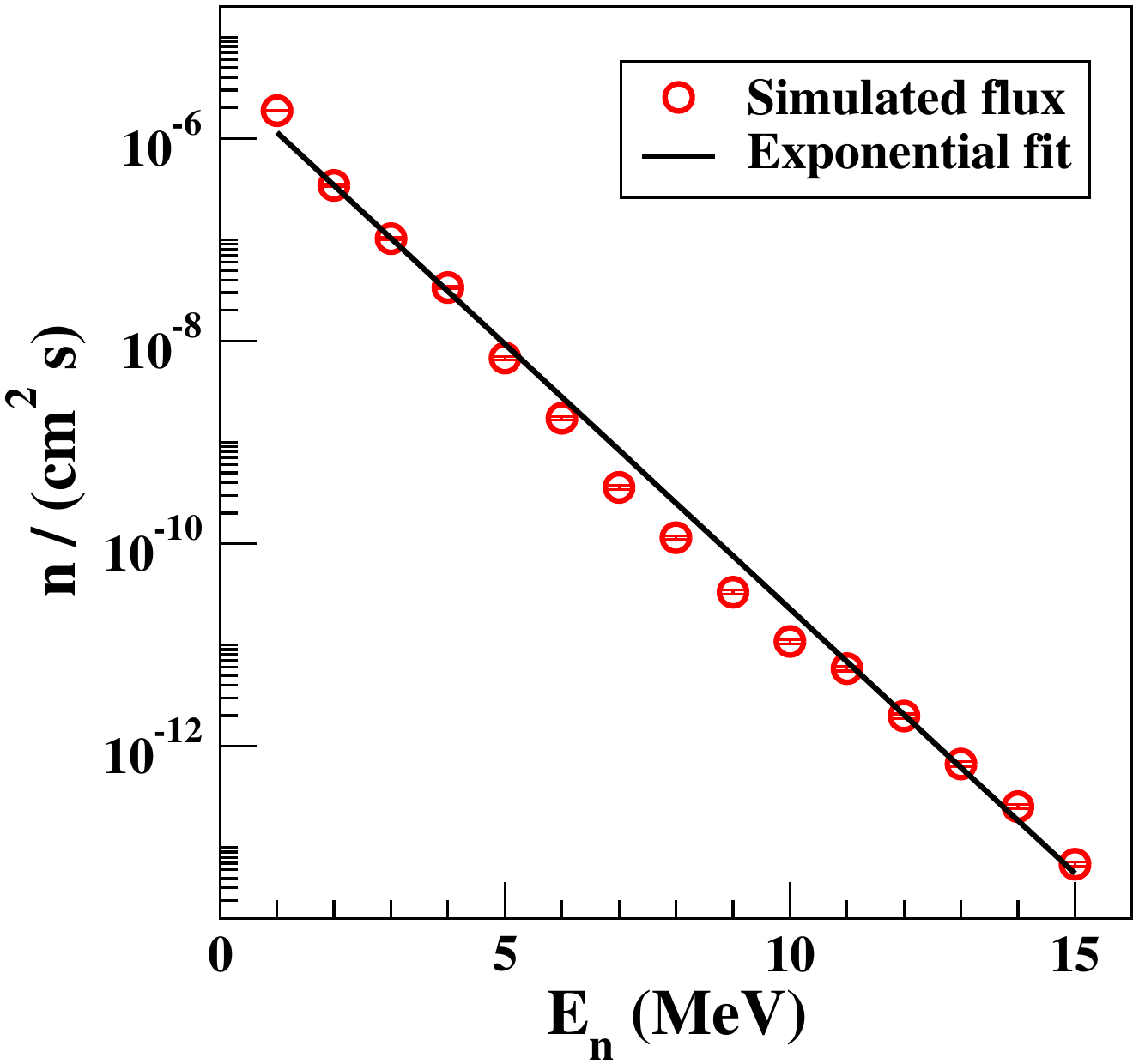} 
\caption[The estimated neutron flux at the center of an underground tunnel at the INO site]{\label{fig:spectra} The estimated neutron flux at the center of the tunnel at the INO site.} 
\end{center}
\end{figure}

Since the neutron source in the rock material is very weak (U, Th in ppb levels), large scale simulations are necessary to estimate the neutron flux. To minimize computation time and statistical errors in simulations, an alternative approach has been employed. Though the INO cavern will have a rock cover of $\sim$ 1 km on all sides, the neutrons will get attenuated very quickly and hence only a finite size rock element will contribute to the neutron flux in the cavern. The size of the rock element was optimized using GEANT4-based MC simulations employing G4NDL4.2 neutron cross-section library. The enhancement in the low energy neutron yield due to scattering in the surrounding rock material has been taken into account. It was observed that the neutron flux at any point on the cavern surface can be computed considering the contribution from a cylindrical rock element of size $\phi=L=140$ cm. 
The neutron flux at the center of the cavern is estimated assuming a cylindrical tunnel of radius $r=2$ m and $L=12$ m and is shown in Figure~\ref{fig:spectra}.
The energy integrated neutron flux from the BWH rock activity thus obtained is $3.2\times10^6\rm~n~cm^{-2}~s^{-1}$. 

The neutron background from rock ($E_n \leq 15$ MeV) can be reduced with passive shield of hydrogen rich materials.
In hydrogenous materials neutron capture by proton releases a $\gamma$-ray ($E_{\gamma}\sim$ 2224.573 keV) very close to the $ Q_{\beta\beta} $ ($ ^{124} $Sn). Simulations indicate that a two layer composite shield of borated paraffin (20 cm) and Pb (5 cm) will be adequate to attenuate the neutron flux ($E_n\leq 15$ MeV) by a factor of $\sim10^{6}$ at the target site.

%%%%%%%%%%%%%%%%%%%%%%%%%%%%%%%%%%%%%%%%%%%%%%%%%%%%%%%%%%%%%%%%%%%%%%%%%
%%
%%   use this format to include an .eps figure into your paper
%%

%%%%%%%%%%%%%%%%%%%%%%%%%%%%%%%%%%%%%%%%%%%%%%%%%%%%%%%%%%%%%%%%%%%%%%%%%%%

\section{Summary}
Radiation background studies pertaining to $0\nu\beta\beta$ decay in $^{124}$Sn have been carried out. A TiLES setup has been installed at TIFR for this purpose. Neutron-induced background is studied in the TIN.TIN detector materials using fast neutron activation technique.
The neutron flux ($E_n\leq15$ MeV) resulting from SF and ($\alpha, n$) interactions for the rock in the INO cavern is estimated using MC simulations. A two layer composite shield of borated paraffin (20 cm) + Pb (5 cm) is proposed for the reduction of neutron flux.

\bigskip
\section{Acknowledgments}
We are grateful to the PLF staff, the INO and TIN.TIN collaborations.

\end{document}